\documentclass[showpacs,preprintnumbers]{revtex4}
\def\be{\begin{equation}}
\def\ee{\end{equation}}
\def\bea{\begin{eqnarray}}
\def\eea{\end{eqnarray}}

\def\d{{\rm d}}

\def\3vd{\rangle{\hspace{-0.18em}\longrightarrow}{\hspace{-0.65em}^\mid}}
\def\4vd{\rangle{\hspace{-0.15em}\longrightarrow}}

\def\2d{\dot{2}}
\def\1d{\dot{1}}
\def\3d{\dot{3}}
\def\e{\epsilon}

\def\kk{\kappa}

\def\m{\mu}
\def\n{\nu}

\def\1q{Q+1}
\def\2q{Q+\kk(1,2)}
\def\3q{Q+\kk(1,3)}

\def\p{\partial}
\def\a{\alpha}
\def\b{\beta}
\def\t{\theta}

\usepackage{graphicx}
\usepackage{dcolumn}
\usepackage{bm}
\usepackage{amsmath}
\usepackage{amssymb}
\usepackage{epsfig}

\begin{document}

\title{
The photo-balls and static solutions in NCQED with time attended}
\author{Abolfazl\ Jafari}
\affiliation{Department of Physics, Faculty of Science,
Shahrekord University, P. O. Box 115, Shahrekord, Iran.\\
jafari-ab@sci.sku.ac.ir}
\date{\today }

\begin{abstract}

We drive the potential of photon interaction from Feynman diagrams amplitudes,
and we show that the photo-balls,
can be produced in
noncommutative electrodynamics with time attended
but
for the static and localized fields, the static solutions (the lumps) can not be exited.
\end{abstract}


\maketitle


\section{Introduction}

Ordinary
noncommutative theories are based on an antisymmetric quantity
of rank two, so it does not exhibit the Lorentz symmetry and it does not make much sense to look
for noncommutative theories invariant by general coordinates \cite{Neto}. But, many people believe that this
problem can be solved by Hopf algebra. This hope has led to further study of noncommutative dynamics.
These studies revealed some peculiar features of noncommutative quantum models.
Much attention has been paid also to quantum field theories on noncommutative space time, in particular
noncommutative Yang-Mills theory as well as noncommutative QED.

The aim of this paper is to study another aspect of the noncommutativity framework adapted to the source-free
static solutions of noncommutative Maxwell equations and extracting photo-balls. As well
known, the Maxwell's four laws describe the
evolution in time and space of the electric and magnetic fields and the photo-balls are the bound states of photon interactions.
%Recently, many people attended to a new approach leading to noncommutative electrodynamics
%with many details \cite{jaf,nekrasov,jab,weise,jafari,jafari2}.
In this paper we will study of values of \textbf{E} and \textbf{B} and we show that
there are no static solutions for source-free noncommutative $U_{\star}(1)$
in the case of $\t^{0i}\neq0$ only.
%In this case, we will remove all $\star$'s related to the metric tensor, because we
%do not search the momentum of the metric.
In Ref.\cite{Deser} S. Deser presents the static solutions in source-free
Yang-Mills theory are forbidden. His work is on the nonabelian electrodynamics in the commutative space time and we think
this idea can not be generalized to noncommutative electrodynamics, comprehensively.
\section{The presentation of photo-balls with time attended}

Interestingly one finds the situation
very reminiscent to that of non-Abelian gauge theories, and then the question is whether
there are some kinds of bound states in analogy with glue-balls of QCD, here might be
called "photo-ball"s.
In previous work Ref.\cite{jafari}, in ncqed with space noncommutativity case, we have shown the photo-balls
%as a bound state of photon interactions,
can be excited.
Now, we show that
in contrast to QED
on ordinary space time, noncommutative QED with time attended
$[x^\m,x^\n]_\star=\imath\delta^{\n0}\theta^{\m\n}$
is
involved by direct interactions between photons.
This work is based on the potential model which this model is a Furrier transformation of Feynman diagrams amplitudes.

We consider the possibility that photons of noncommutative QED can make
bound states on potential model.
The basic ingredient of potential model is that the self-interacting
massless gauge particles may get mass by inclusion non-perturbative effects.

There are two related issues when we are considering the effective gauge theory of
constituent photons as massive vector particles. First, it is known that the gauge symmetry
is lost via the mass term, and the second, massive gauge theories are known to be
perturbatively non-renormalizable. Here we remember of given comments on these issues in Ref.\cite{jafari2}. The
non-renormalizability of massive gauge theories for QCD and NCQED is same and it is
under this assumption that the mass
in the theory appears as a fixed parameter, surviving at large momentum. In fact the
insufficient decrease of propagator of a massive vector particle at large momentum, due
to simple power counting, suggests that the theory can not be renormalizable. But the
situation might be different in a theory with constituent mass. At very large momentum,
where coupling constant is small due to asymptotic freedom, the perturbation is valid and
gluons or photons appear as massless particles. So the mass of constituent gluon or photon, which are generated
dynamically, depends on momentum and vanish at large momentum.
In a theory for
gluons or photons, it is argued that if one can keep the dependence of constituent mass on momentum,
which of course is possible only by including the non-perturbative effects, the theory
may appear to be non-perturbatively renormalizable.
Although the argument above is for a model involved by dynamically generated mass,
due to lack of a systematic treatment of non-perturbative effects, much can be learned
via a kinematical description of gluon or photon mass, it is to assume mass as a fix parameter,
though the problem still remains with local gauge symmetry.
%Many details of these issues are given in ref \cite{jafari2}.

Following the procedure developed for NCQED case, we insert a mass term to
noncommutative QED. As described this is done by introducing an extra
scalar field, so the extra scalars do not appear as external legs of diagrams,
but the situation is even simpler as far as one considers just the tree diagrams,
in which one can ignore the scalars. There are 3 and 4 photon vertices.

%via Ref. \cite{jafari} the vertex functions are
%
%\bea
%\Gamma^{\mu_1\mu_2\mu_3}_{k_1,k_2,k_3}&=&
%-2e\sin\!\big(\frac{k_{1}\wedge k_{2}}{2}\big)
%\Big[(k_{1}-k_{2})^{\mu_{3}}g^{\mu_{1}\mu_{2}}\nonumber\\&+&
%(k_{2}-k_{3})^{\mu_{1}}g^{\mu_{2}\mu_{3}}+ (k_{3}-k_{1})^{\mu_{2}}g^{\mu_{3}\mu_{1}}\Big],\nonumber\\&&
%\eea
%
%and
%
%\bea &&
%\Gamma^{\mu_1\mu_2\mu_3\mu_4}_{k_1,k_2,k_3,k_4}=-4\imath e^{2}\nonumber\\&&
%\Big[\sin\!\big(\frac{k_{1}\wedge k_{2}}{2}\big)\sin\!\big(\frac{k_{3}\wedge
%k_{4}}{2}\big)\big(g^{\mu_{1}\mu_{3}}g^{\mu_{2}\mu_{4}}\!\!-\!g^{\mu_{1}\mu_{4}}g^{\mu_{2}\mu_{3}}\!\big)
%\nonumber\\
%&&+\sin\!\big(\frac{k_{3}\wedge k_{1}}{2}\big)\sin\!\big(\frac{k_{2}\wedge
%k_{4}}{2}\big)\big(g^{\mu_{1}\mu_{4}}g^{\mu_{2}\mu_{3}}\!\!-\!g^{\mu_{1}\mu_{2}}g^{\mu_{3}\mu_{4}}\!\big)
%\nonumber\\
%&&+\sin\!\big(\frac{k_{1}\wedge k_{4}}{2}\big)\sin\!\big(\frac{k_{2}\wedge
%k_{3}}{2}\big)\big(g^{\mu_{1}\mu_{2}}g^{\mu_{3}\mu_{4}}\!\!-\!g^{\mu_{1}\mu_{3}}g^{\mu_{2}\mu_{4}}\!\big)\Big],
%nonumber\\&&
%\eea
%
%,
%and
%the momenta and indices are given in Fig. 2.
In the each vertex energy momentum conservation
should be understood and
we work in the non-relativistic limit, namely \cite{jafari}
$
p^{\mu}=\big(m+\frac{\mathbf{p^{2}}}{2m},\mathbf{p}\big),
$
and for polarization vector
$\epsilon ^{\mu}=\big(\frac{\mathbf{p}\cdot
\mathbf{e}}{m},\mathbf{e}+\frac{\mathbf{p}\cdot\mathbf{e}}{2m^{2}}\mathbf{p}\big),
$
where $\textbf{e}$ is a 3-vector satisfying $\mathbf{e}^*
\cdot \mathbf{e}=1$ and from Lorentz gauge-fixing condition, we have
%
%\bea
$p\cdot \epsilon=p^{\mu} \epsilon_{\mu}=0$
%\eea
%
Although, there are four diagrams at tree level, those coming from s-, t-, u- and seagull channels.
When extracting the potential, by the properly symmetrized wave function for identical particle
systems, the ``exchange" or ``symmetry" diagrams are automatically
taken care of, causing that only one of t- and u- channels'
contributions should be added to others' contributions.
We define the vector $\bm{\theta}$ based on $p_0\theta^{0i}$ due to
$q_0=0$ and $p_0=m+\frac{\textbf{P}^2}{2m}$.

By this vector, for $b=(0,\textbf{b})$, we can write the $\wedge$-product as
\bea
a \wedge b=\theta_{\a\b}a^{\a}b^{\b} =\bm{\theta}\cdot \bm{b}
\eea
Where $(\bm{\theta})^i=p_0\theta^{0i}$.
There are two diagrams
in seagull channel, one gives the contribution ${\cal M}^{\rm s.g.(1)}_{fi}\propto 1+3\rightarrow2+4$
and the other, ${\cal M}^{\rm s.g.(2)}_{fi}$, is obtained with replacements $3\leftrightarrows 4$.
We continue in the center-of-mass
frame.
By referring to Ref.\cite{jafari2}, and for the small noncommutativity parameter,we obtain
\bea
\imath {\cal M}^{\rm s.g (1)}_{fi}&=&8\imath e^{2}
\sin\!\big(\frac{p_{1}\wedge p_{3}}{2}\big)
\sin\!\big(\frac{p_{2}\wedge p_{4}}{2}\big)
\big(\cdots\big)\propto \big(\theta p_0\big)^2 \bf{p}^2
\eea
and
\bea
\imath{\cal M}^{\rm t}_{fi}&=&-4\imath e^2\frac{\sin^2\!\big(\frac{p\wedge
q}{2}\big)}{\mathbf{q}^{2}+m^{2}}   \big[4m^{2}+3\mathbf{q}^{2}
-2\mathbf{S}^{2}\mathbf{q}^{2}+2(\mathbf{S}\cdot
\mathbf{q})^{2}+6i\mathbf{S} \cdot (\mathbf{q} \times
\mathbf{p})\big]+O(\mathbf{p}^2)
\eea
Which the seagull channel is in order of $\mathbf{p}^2$ and we can ignore it.
By replacing the non-relativistic limit of $\epsilon$'s, we see that,
even without considering coefficient involving sin($\cdots$), the leading order contribution of
s-channel is order of $|\mathbf{p}|^{2}\ll m^2$, that we can ignore it in comparison with the
zeroth orders.
This observation is exactly as the same as that happens in the QCD case.
%\cite{corn-soni,hou-luo-wong,hou-wong}.
And a similar one for ${\cal M}^{\rm s.g (2)}_{fi}$ by replacements
$3\leftrightarrows 4$ occurred .
This observation is different
from that for QCD glue-balls, for them the contribution of seagull channel
is in zeroth order of momentum and thus should be kept.\\

\textbf{The effective potential between photons}
%\end{center}

For the t-channel contribution with help of above equation we have
\bea
\imath{\cal M}^{\rm t}_{fi}=-4\imath e^2
\frac{\sin^2\!\big(\frac{1}{2}\mathbf{q}\cdot\bm{\theta}\big)}{\mathbf{q}^{2}+m^{2}}
\Upsilon (\mathbf{q}),
\eea
where
\bea
\Upsilon (\mathbf{q})=4m^{2}+3\mathbf{q}^{2}-2\mathbf{S}^{2}\mathbf{q}^{2}
+2 (\mathbf{S}\cdot\mathbf{q} )^{2}+6\imath\mathbf{S} \cdot  (\mathbf{q} \times
\mathbf{p} ),\nonumber\\&&
\eea
By using the total amplitude, the potential can be written
%deduced using (\ref{potential})
%
\bea
V_{2\gamma}(\mathbf{r})&=&\int
\frac{{\rm d}^{3}q}{8\pi^{3}}\frac{\imath\e^{\imath\mathbf{q}\cdot\mathbf{r}}}
{4\sqrt{E_{1}E_{2}E_{3}E_{4}}}\,\imath{\cal M}_{fi}
\nonumber\\
\eea
Now, we define $U(R)=\int\frac{\d^{3}q}{8\pi^{3}}
\frac{\e^{i\mathbf{q}\cdot\mathbf{R}}}{\mathbf{q}^{2}+m^{2}}$ and by replacing $\mathbf{q}\rightarrow-i\bm{\nabla}$
also we keep the $V_{2\gamma}(\mathbf{r})$ up to first power of $\bm{\theta}^2$ and we ignore from higher power of $(\bm{\theta}^4)$ or we have
\bea
V_{2\gamma}(\mathbf{r})=\frac{e^{2}}{4m^{2}}
\Upsilon(-i \bm{\nabla})
\Big[2\, U(r)-U(r_+)-U(r_-)\Big],
\eea
with $\mathbf{r}_\pm=\mathbf{r}\pm\bm{\theta}$.
We mention that,  only for $\bm{\theta}=0$ the potential vanishes and this is different from space noncommutativity QED where
in the space noncommutativity qed this happens when $\bm{\theta}=0$, $\ \ \ \textbf{p}=0$ and $\textbf{p}\parallel\bm{\theta}$.
It is reasonable to see the behavior of potential for small noncommutativity parameter,
%defined here by $\theta\ll r$ and $\theta m\ll 1$. In this limit,
the first surviving terms are given by
\bea
V_{2\gamma}(\mathbf{r})=-\frac{e^{2}}{4m^{2}}\Upsilon(-\imath \bm{\nabla})
(\bm{\theta}\cdot\bm{\nabla})^2 U(r)
+O(\theta^4)
\eea
Recalling that for a function $f(r)$, $\partial_i f(r)=x_i \nabla_r f$, with
$\nabla_r=r^{-1}\partial_r$, and using $(\mathbf{p}\times\mathbf{S})\cdot\mathbf{r}=(\mathbf{r}\times\mathbf{p})\cdot\mathbf{S}=\mathbf{L}\cdot\mathbf{S},\\
\nabla^2 U(r)= m^2 U(r) - \delta(\mathbf{r})$
with $\mathbf{L}$ as the total angular momentum, we get the expression for potential
\bea\label{pot-1} &&
V_{2\gamma}(\mathbf{r})\!\!\!=\ -\frac{e^{2}}{4m^{2}}\Bigg\{
m^{2}\big(1+2S^2\big)\bigg[(\bm{\t}\cdot\bm{\t})^{2}\nabla_r +\big(\bm{\theta}\cdot\mathbf{r}\big)^2\nabla_r\nabla_r\bigg]
-2\bigg[\Big[S^2(\bm{\t}\cdot\bm{\t})^2+2\big(\bm{\theta}\cdot\mathbf{S}\big)^{2}\Big] \nabla_r\nabla_r
+\big(\bm{\theta}\cdot\mathbf{r}\big)^{2}\nonumber\\
&&\big(\mathbf{S}\cdot\mathbf{r}\big)^{2}\nabla_r\nabla_r\nabla_r\nabla_r
+\Big[4\big(\bm{\theta}\cdot \mathbf{S}\big)\big(\bm{\theta}\cdot\mathbf{r}\big)\big(\mathbf{S}\cdot\mathbf{r}\big)
+(\bm{\t}\cdot\bm{\t})^{2}\big(\mathbf{S}\cdot\mathbf{r}\big)^{2}+S^2\big(\bm{\theta}\cdot \mathbf{r}\big)^2\Big]\nabla_r\nabla_r
\nabla_r\bigg]
+6\bigg[\Big[2\big(\bm{\theta}\cdot\mathbf{r}\big)\big(\mathbf{p}\times\mathbf{S}\big)\cdot\bm{\theta}\nonumber\\&&+(\bm{\t}\cdot\bm{\t})^{2}\big(\mathbf{L}
\cdot\mathbf{S}\big)\Big]
\nabla_r\nabla_r
+\big(\bm{\theta}\cdot\mathbf{r}\big)^{2}\big(\mathbf{L}\cdot\mathbf{S}\big)
\nabla_r\nabla_r\nabla_r\bigg]\Bigg\}\;U(r)\nonumber\\
&&
+{\rm D.D.} + O(\t^4).\nonumber\\&&
\eea
Where $S\ \equiv\ |\mathbf{S}|$, and D.D. is for the distributional derivatives of the function $U(r)$, containing
$\delta$-function and its derivatives.
We make comments on the potential given by Eq. (\ref{pot-1}). First we mention that due to $\mathbf{r}$'s in the inner products,
the effective lowest power is $r^{-5}$. Second,  the strength of the potential, through the definition of
$\bm{\theta}$, depends on mass and momentum. Third, the spin-independent
part of the potential, $S=0$, we have
\bea\label{pot-2} &&
V_{2\gamma}^{S=0}(\mathbf{r})=-\frac{e^{2}}{4}\frac{\e^{-mr}}{4\pi}
\bigg[-(\bm{\theta}
%\cdot\bm{\theta}
)^{2}\frac{mr+1}{r^{3}}+(\bm{\theta}\cdot
\mathbf{\hat{r}})^2\frac {m^{2}r^{2} +3mr+3}{r^{3}}\bigg],\nonumber\\&&
\eea
this potential is not included the photons momentums and this is in contrast of the space noncommutativity potential with $S=0$. The
following modes can be considered
\bea
\bm{\theta}=p_0\theta^{03}\hat{z}
\eea
so we have $\bm{\theta}\cdot\hat{\bm{r}}=p_0\theta^{03}\cos{\theta}$ so $V_{2\gamma}^{S=0}(r,\t)$ is not a central force! but the $p_\phi$ is still
constant of motion
so the motion stays at plane, the meaning of this statement is for a specified amount of $\t$,
according to the behavior of potential, the bound state(s) can be still exist.
For the general $\bm{\theta}$ the potential is $V_{2\gamma}^{S=0}(r,\t,\phi)$ which there are not constant of motion so this case is not
interesting.

\section{Static solutions and photo-balls}

We now turn to the case of noncommutative $U_\star(1)$ theory on the $D$-dimensional Minkowski space time
with noncommutative coordinates.
%in the case of
The $U_\star(1)$ action is
\bea \label{action}
S=-\frac{1}{4}\int d^d x\  (F_{\m\n}\star F_{\a\b})\eta^{\m\a}\eta^{\n\b}
\eea
where $F^{\a\b}=\p^{\a}A^{\b}-\p^{\b} A^{\a}-\imath e [A^{\a},\ A^{\b}]_\star$ denotes the strength of the
noncommutative $U_{\star}(1)$ gauge fields. In space time noncommutativity with condition of
$[\hat{x}^i,\ \hat{x}^j]_\star=\imath\t^{ij}$
%we can(must) remove all $\star$'s related to the metric tensor, because we are not searching for the momentum of metric.
In the standard $U_\star(1)$ theory the canonical energy momentum tensor is
\bea\label{energy momentum}
4T^{\m\n}_\star=-2\{F^{\m\a},\ F^{\ \ \n}_{\a}\}_{\star}-\eta^{\m\n}(F^{\a\b}\star F_{\a\b}),
\eea
%
%and of course,
%
%\bea
%4\mathbf{T}^{\m}_{\star\n}=2\{F^{\m}_{\a},\ F_{\n}^{\a}\}_{\star}-\eta^{\m}_{\n}F^{\a}_{\b}\star F_{\a}^{\b}
%\eea
%
where
\bea\label{motion equation}
D_{\m}\star T^{\m\n}_\star=\p_{\m}T^{\m\n}_\star-\imath e[A_{\m},\ T^{\m\n}_\star]_{\star}=0
\eea
Here, we show that for the case $d=4$ and $\t^{0i}\neq0$
there are no static solutions for electric fields ($F_{i0}=E_i$) to self interacting models of $U_{\star}(1)$ type.

\textbf{Lemma}: \textit{In the noncommutativity QED, the static solutions of electric fields ($F_{0i}$) are absence.}

\textbf{proof}:
The time independent solutions of the gauge fields are chosen so the electric
field reduces to $F_{0i}=\p_0 A_i-\p_i A_0-\imath e [A_0,\ A_i]_\star\mid_{for\ static\ solutions}=-D_i\star A_0$
where $D_i=\p_i-\imath e[A_i,\ ]_\star$ it follows that
$D^i\star F_{0i}=0$ so we have $\int d^{d-1}x A_0\star( D^i\star F_{0i})=boundary\ value-\int d^{d-1}x (D^i\star A_0)\star F_{0i}=
-\int d^{d-1}x F^{2}_{0i}=0$ and consequently that for any values of '$d$' we have $F_{0i}=0$ so it's result is that
absence of the electric fields. In this case, there are no sentences about the magnetic fields.

\textbf{Lemma}: \textit{In the time noncommutativity QED, the static solutions are absence, expect for d=5.}

\textbf{proof}:

In this case of noncommutativity, $[\hat{x}^\m,\ \hat{x}^\n]_\star=\imath\delta^{\n0}\t^{\m\n}$
the all details in Ref.\cite{Deser} are correct or in this case
the previous lemma is still valid and
the stress tensor for a $U_{\star}(1)$ field has following components
\bea
\int d^{d-1}x\ T^{\m}_{\star\m}=\int d^{d-1}x\ \frac{1}{4}(4-d)F^{\a\b}F_{\a\b},
\eea
then
$\int d^{d-1}x\ T^{0}_{0}=\int d^{d-1}x\ \frac{1}{2}(F^2_{0i}+\frac{1}{2}F^2_{ij})$ where $F^2=F^{\m\n}F_{\n\m}$
now, compactness of gauge group indeed to $F^2_{0i}$ and $F^2_{ij}$ be positive.
We assume that the fields will be
vanished on space boundaries because all of $F^{\m\n}$ to fall of faster than $\mid \vec{r}\mid^{-\frac{1}{2}(d-1)}$, so
\bea
\int\ d^{d-1}y\ \textbf{T}^{j}_{j}(y)=0
\eea
The vanishing of the integration implies that
\bea
\int d^{d-1}y\ T^{i}_{i}=\int d^{d-1}y\ \frac{1}{2}((d-3)F^2_{0i}+\frac{1}{2}(5-d)F^2_{ij})=0\nonumber\\&&
\label{4}
\eea
For $d=4$, $F_{0i}$ and $F_{ij}$ must all vanish. For $d>5$ we learn nothing further above
from Eq. (\ref{4}).

For earlier case $[\hat{x}^\m,\ \hat{x}^\n]_\star=\imath\delta^{\n0}\t^{\m\n}$ the photon self interaction will be removed
because the field strength tensor becomes $F_{\m\n}=\p_\m A_\n -\p_\n A_\m=f_{\m\n}$ and there is no self interaction, so we can write

\textbf{Lemma}: \textit{In the noncommutativity with time attended, the photo-balls can not be exited.}

\section{Discussion}

In this work by using the electrodynamics in noncommutative space time
, we drive the potential of $\gamma-\gamma$ interaction based on Furrier transformation of the Feynman diagrams amplitudes.
For special cases we show that the photo-balls can be excited but it can not be produced in noncommutativity with time attended.
Also we show that
the vanishing of self-stress for static systems excludes finite energy time-independent solutions of source-free
$U_{\star}(1)$ theory in (3+1) dimensions. This implies that static solutions in the case of $\t^{0i}\neq0$,
for non-commutative electromagnetic
fields are forbidden.
For the case of spatial noncommutativity, we show that just the electric fields is absence and
the magnetic fields is not.
\section{Acknowledgments}
The author thanks the Shahrekord University for support of this research.
\newline
\\
%\appendix
%\begin{center}
%\subsection{Massive Photon-Photon Scattering Amplitude}

%%%%%%%%%%%%%%%%%%%%%%%%%%%%%%%%%%%%%%%%%%%%%%%%%%


\begin{thebibliography}{99}

\bibitem{Neto} J. Barcelos-Neto, "Noncommutative filed in curved space" and
R. Amorim, J. Barcelos-Neto,
"Remarks on the canonical quantization of noncommutative theories",
J. Phys. {\bf A 34} (2001) 8851-8858.

\bibitem{Deser} S. Deser, "Absence of static solutions in source-free Yang - Mills theory",
Cern - Geneva, ref, {\bf Th.2214}, Cern

\bibitem{jafari} Amir H. Fatollahi, Abolfazl Jafari, "On The Bound States Of Photons In Noncommutative Quantum Electrodynamics"
Eur. Phys. J. {\bf C46} (2006) 235-245.

\bibitem{jafari2} Abolfazl Jafari,
"Amplitudes dependent of helicity confugurations QED", Eur. Phy. J. {\bf C 70} (2010) 1131-1143, hep-th/0609083.

\end{thebibliography}
\end{document}